\documentstyle[preprint,aps]{revtex}
\preprint {IMSc-94/01}
\begin{document}
\draft
\title{
Haldane exclusion statistics and second virial coefficient
}

\author{ M. V. N. Murthy
 and R. Shankar}

\address
{Institute of Mathematical Sciences, Madras 600 113, India.\\
}
\date{\today}
\maketitle
\begin{abstract}
We show that Haldanes new definition of statistics, when generalised to
infinite dimensional Hilbert spaces, is equal to the high temperature
limit of the second virial coefficient. We thus show that this exclusion
statistics parameter, g , of anyons is non-trivial and is completely
determined by its exchange statistics parameter $\alpha$. We also
compute g for quasiparticles in the Luttinger model and show that it is
equal to $\alpha$.
\end{abstract}


\narrowtext
There has been much interest in the physics of anyons in the past few
years. Anyons are particles whose many particle wavefunctions pickup a
phase $e^{i\alpha \pi}$ under the exchange of the positions of any two
particles\cite{leinaas}. Arbitary values of $\alpha$ are allowed by the
configuration space topology of many
particle systems in one and two spatial dimensions.  In particular $\alpha=0$
corresponds to completely symmetric wavefunctions (bosons) and
$\alpha=1$ corresponds to antisymmetric wave functions (fermions).
The parameter $\alpha$ is traditionally called the statistics of the
particle, since for $\alpha =1$, the antisymmetry of the wavefunction
implies the Pauli exclusion principle.  This has non-trivial
consequences for the counting of states in the many particle system
and hence on the statistical mechanics.

The effect of the exchange phase on the exclusion principle for
arbitrary $\alpha$ was largely unexplored until recently. In a seminal
paper\cite{haldane}, Haldane has proposed an alternate definition of
statistics based on
a generalised exclusion principle.  His definition is for particles
with finite dimensional Hilbert spaces but can, in principle,  lead to
the existence of fractional
statistics in arbitrary spatial dimensions. The parameter governing
fractional statistics in this case is defined by $g =
-\frac{(d_{N+\Delta N} -d_N)}{\Delta N}$, where N is the number of
particles and $d_N$ is the dimension of the one particle subspace in N
- particle Hilbert space. By definition, therefore, $g=0$ corresponds to
bosons and $g=1$ to fermions. Haldane applied this
definition to two cases:  Quasi-particles in fractional quantum Hall
effect (FQHE) systems and spinons in quantum antiferromagnets. In both
cases he argued that the exclusion statistics parameter, $g$ is equal to
the exchange statistics parameter $\alpha$.   Johnson and
Canright\cite{johncan}
have tested this prediction numerically for FQHE systems and find this
 to be true with
the modification that for quasiparticles, $g=2-\frac{1}{3}$ where as
$\alpha = -\frac{1}{3}$. For the case of anyon gas, Haldane argued
that since the coupling of particles to the Chern-Simons gauge field
cannot change the dimension of the Hilbert space, the exclusion
statistics of anyons is that of the original hard-core particles, ie.,
$g=1$.

In this paper we first generalize Haldane's definition to the case
where the Hilbert space of the particles is infinite dimensional.  We
then show that in general the exclusion statistics parameter $g$ is
proportional to the dimensionless second virial coefficient provided
the system admits a virial expansion of the equation of state in the
high temperature limit.  We apply this result to the case of
anyon gas and find that $g = 2\alpha -\alpha^2$, when the exchange
statistics parameter $\alpha$ , is chosen to be in the range $0 \le
\alpha \le 2$.  We then analyse the case of anyons in a magnetic
field, confined to the first Landau level.  We obtain $g=\alpha$ for
this case.   We also apply
our result to the case of quasiparticles in a one component Luttinger
liquid.  These quasiparticles also have a nontrivial exchange phase
due to the braiding properties of the vertex operators that create
them.  Here also we find that $g=\alpha$,  the spinons are a
special case of this when $\alpha = 1/2$. Note for quasiparticles with
exchange statistics $-1/3$,
we have to take $\alpha = 2-1/3$.  Thus our definition and calculation
is consistent with the numerical findings\cite{johncan}.

The dimension of the Hilbert space of N identical
particles when $g$ is the statistics parameter is given
 by,\cite{haldane,johncan}.
\begin{equation}
D_N(g) = \frac{(d+(1-g)(N-1))!}{N! (d-1-g(N-1))!},
\end{equation}
where d is the dimension of the single particle space.  It is easy to
see that $g=0 (1)$ yields the dimension of the bosonic(fermionic)
Hilbert space.  It is now straight forward to extract the statistics
parameter $g$ from eq.(1) in the limit where the single particle
dimension $d \rightarrow \infty$, we obtain
\begin{equation}
\frac{1}{2} -g = \lim_{d \rightarrow \infty }
\frac{d}{N(N-1)} \left[ N! \frac{D_N(g)}{d^N} -1\right],
\end{equation}
A regulated definition of the dimension of the Hilbert space is given by
the corresponding N-particle partition function since we have,
$ D_N = \lim_{\beta \rightarrow 0} Z_N =
\lim_{\beta \rightarrow 0} Tr (e^{-\beta H_N}), $
where $\beta$ is the inverse temperature and $H_N$ denotes the
N-particle Hamiltonian.  Therefore, for dealing with infinite
dimensional Hilbert spaces, we propose a generalisation of Haldane's
definition,
\begin{equation}
\frac{1}{2} -g = \lim_{\beta \rightarrow 0 }
\frac{C Z_1}{N(N-1)} \left[ N! \frac{Z_N}{Z_1^N} -1\right],
\end{equation}
where C is an overall constant of proportionality which we fix next.
The definition of $g$ given by eq.(3) forms the basis for the rest of
the calculations presented in this paper.

To test the utility of eq.(3), we first consider the case of bosons
and fermions in $d_s$ space dimensions.  For convenience we confine all
the N-particles in an oscillator potential of the form $V(\vec r_1,
\vec r_2,...,\vec r_N) = \frac{1}{2} m \omega^2 \sum_{i=1}^{\infty}
r_i^2$, where $\omega$ is the oscillator frequency and $\vec r_i$ are
single particle coordinates.  The potential here merely acts as a
regulator and oscillator potential which will be used throughout this
paper is only a convenient choice.  The N-particle partition function
is then given by the expansion\cite{myrheim},
\begin{equation}
Z_N^{B,F} (\beta) = \frac{1}{N!}\left[ Z_1^N(\beta) \pm Z_1(2\beta)
Z_1^{N-2} \frac{N(N-1)}{2} + ... \right],
\end{equation}
where the $\pm$ signs refer to bosons and fermions respectively.
Notice that the second term in eq.(4) involves $Z_1(2\beta)$ since the
contribution to this term comes when two particles are in the same
energy level.  Equation (4) is an obvious generalisation of eq.(1)
with a cutoff on the single particle energies.
The single particle partition function is given by,
\begin{equation}
Z_1(\beta) = \frac{ e^ {-\beta \omega d_s/2}}{(1 - e^{-\beta
\omega})^{d_s}}.
\end{equation}
Substituting eq.(5) and eq.(4) in eq.(3) and taking the high
temperature limit we find,
$
\frac{1}{2} -g = \pm \frac{C}{2 (2)^{d_s}}
$
which essentially isolates the contribution of the subleading order
term in eq.(4).  We therefore set $ C = 2^{d_s}$ and immediately
obtain $g=0(1)$ for bosons (fermions). Thus factor $C$  occurs because
the regulated definition cuts off the energy at $2\beta$ when two
particles occupy the same state, however for counting purposes we want to
cutoff all states at the same energy irrespective of their occupation.

We are now in a position to apply the definition of $g$ as generalised
to infinite dimensional Hilbert spaces by eq.(3) to N-particle systems
with interactions. To begin with consider a system of N-interacting
bosons in two space dimensions.  We first show the connection between
eq.(3) and the second virial coefficient in the equation of state of
the system.  The virial expansion for the pressure P of a gas
is by definition a high temperature or low density expansion in terms
of the particle number density $\rho = N/V$, where N is the particle
number and V is the volume (area in two space dimensions). In the
thermodynamic limit $N \rightarrow \infty$ and $V \rightarrow \infty$
when $\rho$ is held fixed\cite{pathria,diptiman}.  The virial
expansion is then given by
\begin{equation}
\beta P = \rho \left[ 1+\sum_{l=1}^{\infty} B_{l+1} (\rho
\lambda^2)^l \right],
\end{equation}
where $\lambda = \sqrt{2\pi\beta/m}$ is the thermal wave length and
the dimensionless $B_l$'s are called virial coefficients which can be
expressed in terms of the partition functions.  For example the second
virial coefficient $B_2$ is given by,
\begin{equation}
B_2 = Z_1 \left[ 1 - 2 \frac {Z_2}{Z_1^2}\right].
\end{equation}
(Note: Conventionally $B_2$ has $V/\lambda^2$ as an overall factor
instead of $Z_1$ when box normalisation is used.) Comparing eq.(7) and
eq.(3) we immediately obtain $g$ for the case N=2,
\begin{equation}
\frac{1}{2} -g = -2B_2.
\end{equation}
This is an exact result for N=2.  Now consider $N\ge 3$.  The higher
virial coefficients are given by,
\begin{equation}
B_3 = 4B_2^2 -2 b_3,~~~~~~B_4 = 9B_2B_3 -16B_2^2 +3b_4.
\end{equation}
In general we have, for $k\ge 3$
\begin{equation}
B_k = F_k(B_2,B_3,...,B_{k-1}) + (-1)^{k-1}(k-1) b_k,
\end{equation}
where $F_k$ are some functions of $B_2,B_3,...,B_{k-1}$ whose explicit form
is not needed here and $b_k$ is given
by
\begin{equation}
b_k = (Z_1)^{k-1} (-1)^{k-1} \sum_{\{m_i\}} (-1)^{(\sum_i m_i -1)}
(\sum_i m_i -1)! \prod_i \frac{(Z_i/Z_1^i)^{m_i}}{m_i !}.
\end{equation}
The summation over $m_i$ is constrained by $\sum_i^k i m_i =k$.  Thus
the virial expansion of the equation of state exists iff all $b_k$'s
are finite.  In order to proceed with the computation of $g$ for $N \ge
3$ , let us assume that for the system under consideration the virial
expansion exists and hence all $B_k$'s are finite.  The high
temperature expansion for the factor $Z_N/Z_1^N $ in eq.(3) is in
general given by
\begin{equation}
\frac{Z_N}{Z_1^N} = \frac{1}{N!} + f_2^{(N)} (\beta \omega)^2 + ...
\end{equation}
where $f_2^{(N)}$ is some function of the interaction strength and may
on N.  In the above equation we have only shown the
expansion up to next to leading order as $\beta \rightarrow 0$; higher
order terms are irrelevent for our purposes.  Using eq.(12) we
immediately see that the leading divergence in $b_k$ as $\beta
\rightarrow 0$ comes from the
overall factor $Z_1^{k-1}$ and is given by $1/(\beta
\omega)^{2(k-1)}$.  For $b_k$ to be finite in the limit $\beta
\rightarrow 0$, we must demand that all terms in the sum involving
powers of
$(Z_i/Z_1^i)$ in eq.(11) up to order $(\beta \omega)^{(2(k-1)-1)}$
must cancel order by order.  This necessarily implies that the
coefficient of $(\beta \omega)^2$ must be set to zero.  Demanding this
we immediately find,
\begin{equation}
f_2^{(N)} = \sum_{n=1}^{N-2}\frac{f^{(N-n)}_2}{n!} (-1)^{(n-1)} =
N(N-1)\frac{f^{(2)}_2}{N!},
\end{equation}
where all $f_2^{(N)}$ are now related to $f^{(2)}_2$ corresponding to
the two particle partition function.  Substituting eq.(12) and eq.(13)
in eq.(3) we obtain for $g$,
\begin{equation}
\frac{1}{2} -g = f^{(2)}_2 = -2 B_2
\end{equation}
We have, therefore, now a relation between $g$ and the second virial
coefficient $B_2$ for a system for which the virial expansion exists.
Notice that in obtaining the result (14) for $g$, it is sufficient to
demand that order $(\beta\omega)^2$ terms cancel in the sum given in
eq.(11).  This is obviously a much weaker condition than demanding that
the virial coefficients be finite.  Nevertheless the whole analysis makes
sense only if the virial expansion is valid.  This then is the main
result of the paper: {\it The Haldane exclusion statistics parameter $g$
is completely determined by the high temperature limit of the second virial
coefficient of a system of interacting particles which admits a virial
expansion in this limit.}

We can now apply this result to some well known systems.  We first
consider anyon gas.  Here the second virial coefficient is well known
and is given by\cite{arovas},
\begin{equation}
B_2 = - \frac{2\alpha^2 - 4\alpha +1}{4},
\end{equation}
where $\alpha=0 $ is bosonic and $\alpha =1$ is fermionic.  Using
eq.(14) we therefore have,
\begin{equation}
g=\alpha(2 -\alpha)
\end{equation}
which has the correct limit for $\alpha=0,1$.  This result would be
true for two anyons.  However, it is not conclusively proved that the
higher virial coefficients are finite in the case of anyon gas.  It
has been proved that  the third virial coefficient is
finite\cite{myrheim,virial}. If indeed all virial
coefficients are finite, then the result given in (16) would be true
for anyon gas in general.  This is an interesting result since the exclusion
statistics parameter $g$ is not the same as the exchange statistics
$\alpha$ unlike the systems considered by Haldane.
It was expected that the particles constituting anyon gas
would be classified as hardcore bosons (hence fermions) since the
coupling of a particle to Chern-Simons gauge
field does not affect
Hilbert space dimensions.  Obviously our result does not bear out this
argument. It is easy to see why. The two anyon spectrum, after taking out
the centre of mass, is given by  $E_{n,l} = \omega (
2n+|l-\alpha|+1)$. When $\alpha$ is nonzero all the energy levels are
shifted up by $\alpha$ ($l\le 0$) or $2-\alpha$ ($l>0$) ( equivalently
$-\alpha$) no matter where the cutoff lies.

Next consider anyons confined in a magnetic field.  Once again $g$ is
given by eq.(16) when the partition function includes the trace over
all Landau levels.  This is easy to see since both the oscillator
frequency $\omega$ and the cyclotron frequency $\omega_c$ act as
regulators and at high temperatures one recovers the anyon gas result.
If however we confine the trace to a single Landau level, the we
obtain $g=\alpha$ in conformity with the results obtained for FQHE
systems.

We will now consider quasiparticles in the Luttinger
model. The low energy physics of this model maps on to
the massless Thirring model.  As is well
known\cite{known}, the model is exactly solvable. We will therefore be
able to compute the RHS of equation(3) directly in this case.
The theory can be written completely in terms of the left(L) and right(R)
currents satisfying the algebra $[J_n^r,J_m^{s\dag}]=n\delta_{mn}\delta_{rs}$,
 where $n,m=1,2,...$ and $r,s=L,R$.  We also have the zero modes
$[\phi_0^r,J_0^s]=i\delta_{rs}$.  The hamiltonian is
\begin{equation}
H = \frac{2\pi v_F}{L}\sum_r \left[ \frac{1}{2} J_{0r}^2 +
\sum_{n=1}^{\infty} J_n^r J_n^{r\dag} \right].
\end{equation}
Quasiparticles are created by the action of the vertex operators
$V_r^{\dag}(x) = : e^{-i q^r\phi^r(x)}:$ on the ground state.  The
$\phi^r(x)$ are the bosonic phase fields given by,
\begin{equation}
\phi^r(x) = \phi_0^r(x) +  \frac{2\pi}{L} \sigma_r \left[ xJ_{0r} +
\sum_{n=1}^{\infty}\left( \frac{e^{-i\sigma_r\frac{2\pi}{L}nx}}{in}
J_n^r + hc\right)\right],
\end{equation}
where $\sigma_{L,R}= \pm 1$.  The periodicity properties of the
compact zero modes $\phi_0^{L,R}$ constrain the allowed values of
$q_r$ to be $q_r = (NR + \sigma_r \frac{M}{2R})$ and R is the radius of
the field $\phi$.   It is a function of the interaction strength.  Our
conventions correspond to $R=1$ for the case of non interacting
fermions.  The constraints on $q_r$ imply that quasiparticles in the
left sector must be created along with quasiparticles in the right
sector (except for special values of R).  However because the
hamiltonians of the two sectors are completely decoupled, it is
consistent to analyse the spectrum of the two sectors independently.
We will therefore focus on the left sector alone.

We first consider the space of one quasiparticle states which we
define to be the space of the states $|x> = V_L^{\dag}(x) |0>$.  These
are not a linearly independent set.  From the form of $V_L^{\dag}(x)$,
it follows that $|x+L> = e^{i\alpha \pi}|x>$, where $\alpha = q_L^2$.
Therefore, we have the expansion $|x> = \sum_{n=1}^{\infty}
e^{ik_nx}|n>$, where $k_n = \frac{2\pi}{L}(n+\frac{\alpha}{2})$.
{}From the form of $V_L^{\dag}(x)|0>$ it follows that $n>0$.  It can be
easily shown that $|n>$ form an orthogonal set of eigenstates of the
Hamiltonian with eigenvalues $E_n^{(1)} = v_F k_n$.  The single
particle partition function is then given by,
\begin{equation}
Z_1 = \frac{e^{-\tilde \beta \alpha /2}}{1-e^{-\tilde \beta}} =
e^{-\tilde \beta \alpha/2} Z_1^B; ~~~~~~~~\tilde \beta =
\frac{2\pi}{L}\beta v_F.
\end{equation}

Next we come to the N quasiparticle states which we define to be the
span of $|\{x_n\}>=\prod_{n=1}^{N} V_L^{\dag}(x_n)|0>$.  From the
fact that $ V_L^{\dag}(x)  V_L^{\dag}(y)  =e^{i\alpha\pi}V_L^{\dag}(y)
V_L^{\dag}(x)$ , it follows that these many quasiparticle states pick
up a phase $e^{i\alpha\pi}$ under the exchange of two of the
coordinates.  We normal order the vertex operators to obtain,
$|\{x_n\}>=\prod_{n>m} \left[
sin\left(\frac{\pi(x_n-x_m)}{L}\right)\right]^{\alpha}:\prod_{n=1} ^{N}
V_L^{\dag}(x_n):|0>$.
Again from the form of the vertex operators it
follows that we have the expansion,
\begin{equation}
|\{x_n\}>=\prod_{n>m}^{N}
\left[ sin\left(\frac{\pi(x_n-x_m)}{L}\right)\right]^{\alpha}
\sum_{\{k_n\}}\phi_N^B(\{k_n\}|\{x_n\})|\{k_n\}>,
\end{equation}
where $ \phi_N^B(\{k_n\}|\{x_n\})$, are symmetrized N particle plane
waves with momenta $\{k_n\}$, where $k_n =
\frac{2\pi}{L}(n+\frac{N\alpha}{2});~~n>0$. The states
$|\{k_n\}>$  can be shown to be eigenstates of the hamiltonian with
energy $E^N(\{k_n\}) = v_F \sum_{n=1}^N k_n$.  We can show that
$|\{k_n\}>$  are a linearly independent set of states.  The states
with the same energy are however not orthogonal.  The N-quasiparticle
partition function is then $Z_N =
e^{-\tilde \beta N^2\alpha/2} Z_N^B$, where $Z_N^B$ is the N-particle bosonic
partition function.  We can now exactly compute $g$ using the high
temperature expansion for $Z_N^B$ and obtain $g=\alpha$.  Thus the
exchange and
exclusion statistics parameters are identical for these models. Note
that the exchange statistics in one dimensions is somewhat arbitary. We
could have changed it by multiplying the vertex operators with suitable
cocycle factors. The exclusion statistics parameter is however
unambigous and unique.
When $R = \frac{1}{\sqrt 2}$, the theory is equivalent to the low energy
physics of the SU(2) symmetric quantum antiferromagnetic chain.  We
then have the spinon excitations with $\alpha = 1/2$.  Thus we recover
Haldane's result that $g=1/2$ for this case.

The above example gives a clear insight into the mechanism of the
phenomenon. What is happening is that the addition of a quasiparticle
causes a phase shift of every other quasiparticle, resulting in an
energy shift of ${\alpha \pi v_{F}} \over L$ per particle . When we count
the dimension of the
single particle space with a fixed (smooth) cutoff, there are $\alpha$
states missing. The important thing here is that the
all single particle levels shift up by the
same amount however high the energy. This is why we get $g$ to be
well defined and nontrivial in the cutoff going to infinity limit.

Exactly the same thing happens in the case of the anyon gas except that
the relevent levels do not all shift by the same amount.  The
$l=0,-2,-4,...$ levels in the two anyon spectrum shift up whereas the
positive $l=2,4,6,...$ levels shift down by an amount $-\alpha$.
Equivalently the later set may be considered as made up of levels
$l=0,2,4,6,...$
with an upward shift given by $2-\alpha$.   This is why $g$ is not
equal to $\alpha$ but
has a nonlinear dependence on it given by the product of the two
energy shifts, namely $\alpha(2-\alpha)$ as though the full $g$ is the
product of $g$'s corresponding to two one component systems.
However the important fact is that g
is completely determined by $\alpha$. When we consider anyons in a
strong magnetic field and restrict ourselves to the first Landau level,
then again we have $l$'s of only one sign. This again results in
$g=\alpha$.

Thus the fundamental character of exclusion statistical interactions
seems to be that they cause scale-invariant phase shifts and hence
scale-invariant energy
shifts. How this is connected to the non-orthogonality of position
eigenstates which is stressed in ref.\cite{haldane} is not clear to
us. We also note
that these scale-invariant phase shifts, if they occur in any model, is
a non-perturbative effect and the model cannot be analysed by
perturbation theory \cite {pwa}.
Thus a non-trivial $g$ implies the inapplicability of standard many-body
perturbation theory, as has been stated in ref\cite{haldane}.

Finally, we have related $g$ to the high temperature limit of a
thermodynamic quantity, which in principle,
is much easier to measure than the exchange phase of quasiparticles in
condensed matter systems. It may therefore be possible to use this
connection to device a realistic experiment to measure the Haldane
exclusion statistics of quasiparticles.

We gratefully acknowledge many interesting discussions with G.Baskaran.

\end{document}